\documentclass[12pt,a4paper]{article}

\usepackage{colortbl}
\usepackage{ascmac}
\usepackage{bm}
\usepackage[dvipdfmx]{graphicx}
\usepackage{amsmath}
\usepackage{amsfonts}
\usepackage{amssymb}

\def\Pexp{\mathop{\rm Pexp}\nolimits}
\def\Plog{\mathop{\rm Plog}\nolimits}
\def\tr{\mathop{\rm tr}\nolimits}
\def\rank{\mathop{\rm rank}\nolimits}

\newcommand{\CC}{\mathbb{C}}

\newcommand{\RR}{\mathbb{R}}
\newcommand{\ol}{\overline}
\newcommand{\wt}{\widetilde}

\allowdisplaybreaks

\begin{document}

\begin{titlepage}
\title{
\vspace{-1.5cm}
\begin{flushright}
{\normalsize TIT/HEP-701\\ May 2024}
\end{flushright}
\vspace{1.5cm}
\LARGE{Inductive calculation of superconformal indices\\ based on giant graviton expansion}}
\author{
Yosuke {\scshape Imamura$^1$\footnote{E-mail: imamura@phys.titech.ac.jp}}
and
Shuichi {\scshape Murayama$^2$\footnote{E-mail: murayamasab3@gmail.com}}
\\
\\
{\itshape $^1$Department of Physics, Tokyo Institute of Technology}, \\ {\itshape Tokyo 152-8551, Japan} \\[2ex]
\itshape $^2$Independent Researcher}

\date{}
\maketitle
\thispagestyle{empty}
\begin{abstract}
We investigate a simple-sum giant graviton expansion of
the superconformal indices of ${\cal N}=2$ superconformal field theories
realized on D3-branes probing $7$-brane backgrounds with
constant axio-dilation field.
The expansion is of self-dual type,
and imposes strong constraints on the indices.
By using the constraints
we determine first few terms in the superconformal indices for
arbitrary rank $N$.
\end{abstract}

\end{titlepage}

\tableofcontents
\section{Introduction}
AdS/CFT correspondence \cite{Maldacena:1997re,Gubser:1998bc,Witten:1998qj} is usually regarded as
a duality between a $d$-dimensional boundary field theory
and a $d+1$ dimensional bulk theory including gravity.
This duality have been extensively studied in the large $N$ limit,
and a lot of evidences have been found.
For example, in the correspondence between the ${\cal N}=4$ $U(N)$ SYM and
the type IIB string theory in $AdS_5\times S^5$,
the superconformal index
\cite{Romelsberger:2005eg,Kinney:2005ej}, which is the quantity we focus on in this paper,
was calculated on the both sides of the correspondence in the large $N$ limit,
and the agreement was confirmed \cite{Kinney:2005ej}.

The correspondence is believed to hold even for
finite $N$.
Formulas giving the finite $N$ superconformal index on the AdS side
have been proposed \cite{Arai:2019xmp,Imamura:2021ytr,Gaiotto:2021xce},
and are called giant graviton expansions.
In the expansions, the finite $N$ corrections are included as
the contributions of giant gravitons \cite{McGreevy:2000cw,Mikhailov:2000ya} 
wrapped around cycles in
the internal manifolds.
They are calculated as the indices of field theories
on giant gravitons.
Therefore, a giant graviton expansion can be regarded as
a set of relations between two classes of field theories:
boundary field theories with different ranks and
theories realized on brane systems
labeled by wrapping numbers.

We consider superconformal theories $T_N$ with rank $N=1,2,3,\ldots$,
and their superconformal indices $I_N$.
The general form of a giant graviton expansion is
\begin{align}
\frac{I_N}{I_\infty}=\sum_{\vec m} x_1^{m_1N}\cdots x_k^{m_kN}F_{m_1,\ldots,m_k}.
\label{multiple}
\end{align}
where $\sum_{\vec m}$ is the summation over $k$ non-negative wrapping numbers $m_1,\ldots, m_k$.
$k$ depends on the theories;
$k=3$ for ${\cal N}=4$ $U(N)$ SYM and orbifold theories \cite{Arai:2019xmp,Arai:2019wgv}, and for more general quiver gauge theories
realized on D3-branes probing toric Calabi-Yau three-folds,
$k$ is the number of the corners of the toric diagram \cite{Arai:2019aou}.
Other examples have been also studied \cite{Arai:2020uwd,Fujiwara:2021xgu,Imamura:2021dya}.

It was found in \cite{Gaiotto:2021xce} that
(\ref{multiple}) reduces to a
simple-sum giant graviton expansion for ${\cal N}=4$ $U(N)$ SYM
by adopting an appropriate expansion scheme.
Namely, if we treat $F_{m_1,\ldots,m_k}$ as a power series with respect to
an appropriate fugacity $\mathfrak{t}$,
$F_{m_1,\cdots,m_k}$ is non-vanishing only when $m_2=\cdots=m_k=0$ and then the multiple-sum expansion
(\ref{multiple}) reduces to a simple-sum over $m:=m_1$.
See also \cite{Imamura:2022aua,Fujiwara:2023bdc} for the mechanism of the reduction to the simple-sum.
The non-trivial contributions $F_{m,0,\ldots,0}$ are
essentially the
superconformal indices of theories $T_m'$ realized on the worldvolume of the stack of $m$ giant gravitons.
More precisely, $F_{m,0,\ldots,0}$ is obtained from the index $I'_m$ of $T_m'$
by a simple variable change of fugacities \cite{Arai:2019xmp,Gaiotto:2021xce}, which is schematically given by
$F_{m,0,\ldots,0}(\mathfrak{t},s,u_i)=I_m'(\mathfrak{t},s^{-1},\wt u_i)$.
As the result, the giant graviton expansion is given by
\begin{align}
\frac{I_N(\mathfrak{t},s,u_i)}{I_\infty(\mathfrak{t},s,u_i)}=\sum_{m=0}^\infty s^{mN}I'_m(\mathfrak{t},s^{-1},\wt u_i),
\label{nonsd}
\end{align}
where $\wt u_i$ is a certain permutation of $u_i$.
With this relation, we can calculate the indices $I_N$ with arbitrary $N$
from the indices of another class of theories $I'_m$ with different ranks $m$.

In general, $T_N$ and $T'_N$ may be different theories.
For example, the indices of M2-brane theories (ABJM theories) and M5-brane theories ($(2,0)$ theories of type $A$)
are related by the expansion in the form (\ref{nonsd}) \cite{Imamura:2022aua}.
More examples of such pairs of four-dimensional field theories related by (\ref{nonsd}) are studied in \cite{Fujiwara:2023bdc}.
Such relations are useful when we know the indices on one side of the relation.

In this work, in contrast, we consider the self-dual situation:
the case that the theories $T_N$ and $T'_N$ appearing on the two sides of the
giant graviton expansion are the same.
In this case the giant graviton expansion is given by
\begin{align}
\frac{I_N(\mathfrak{t},s,u_i)}{I_\infty(\mathfrak{t},s,u_i)}=\sum_{m=0}^\infty s^{mN}I_m(\mathfrak{t},s^{-1},\wt u_i).
\label{sdexpansion}
\end{align}
This is similar to (\ref{nonsd}), but the same indices $I_N$ (with different arguments)
appear on the two sides of the relation.
This is the case for the ${\cal N}=4$ SYM with $U(N)$ gauge group \cite{Gaiotto:2021xce}.
Another example is the ${\cal N}=2$ theory realized on D3-branes probing an O7-plane background \cite{Fujiwara:2023bdc},
which we denote by $D_4[N]$ below.
In this work, we study a class of ${\cal N}=2$ superconformal field theories
including the above two as special cases;
${\cal N}=2$ superconformal field theories
realized on D3-branes probing $7$-brane backgrounds with
constant axio-dilaton field.

(\ref{sdexpansion}) is expected to hold for an arbitrary $N$, and imposes very strong constraints
on the set of functions $I_N$,
and
it is likely that we can determine complete indices form
a small piece of information obtained by some analysis independent of the giant graviton expansion.
The purpose of this work is to discuss to what extent (\ref{sdexpansion}) constrains
the indices and explicitly determine first few terms of them starting from a
partial information.

\section{Strategy}\label{strategy}
If we know the holographic dual description of $T_N$, we can calculate the large $N$ index $I_\infty(\mathfrak{t},s,u_i)$
by the mode analysis of massless fields on the AdS side.
So, let us treat $I_\infty$ as a known function.
Then, 
(\ref{sdexpansion}) gives infinitely many linear relations
among $I_N(\mathfrak{t},s,u_i)$ and $I_m(\mathfrak{t},s^{-1},\wt u_i)$.

As we mentioned in Introduction,
we treat the index as a power series
\begin{align}
I_N(\mathfrak{t},s,u_i)=\sum_{k=0}^\infty \mathfrak{t}^kf_{N,k}(s,\wt u_i)
\label{texpansion}
\end{align}
in an appropriately chosen fugacity $\mathfrak{t}$.
As we will see later, $\mathfrak{t}$ is a fugacity associated with a Cartan generator
of a non-Abelian symmetry, and the charge is quantized to be integer.
This guarantees that the power series in (\ref{texpansion}) includes only integral powers of $\mathfrak{t}$.
(This is not the case for $s$.
The functions $f_{N,k}(s,\wt u_i)$ include fractional powers of $s$.)
The form of the expansion (simple-sum/multiple-sum)
depends on the expansion scheme, and
the simple-sum expansion works only when we first expand functions
with respect to the specific fugacity $\mathfrak{t}$ as in (\ref{texpansion}).

A nice property of the expansion (\ref{texpansion})
is that a truncation at an arbitrary $k=k_*$ gives
a closed set of relations among $f_{N,k}$ with $k\leq k_*$,
and we can discuss inductive determination of $f_{N,k}$ in the order $k=0,1,2,\ldots$.

Let us first consider the truncation at $k_*=0$,
which gives the following relation among $f_{N,0}$:
\begin{align}
\frac{f_{N,0}(s,u_i)}{f_{\infty,0}(s,u_i)}=\sum_{m=0}^\infty s^{mN} f_{m,0}(s^{-1},\wt u_i).
\label{expansion2}
\end{align}
The presence of $s^{-1}$ on the right hand side makes the analysis complicated.
Naive use of series expansion in $s$ does not work because
the expansion coefficients of $f_{N,0}(s)$ and those of $f_{N,0}(s^{-1})$ are not directly related,
and we need to analytically continue the functions around $s=0$ to those around $s=\infty$.

Fortunately, we know the analytic forms of these functions.
As we will discuss below, the functions are nothing but the Coulomb branch index.
The theory $T_N$ we discuss is defined as the theory on a stack of $N$ D3-branes and $N$ independent
Coulomb branch operators exist.
The Coulomb branch index depends only on $s$, and
we denote $f_{N,0}(s,u_i)$ simply by $f_N(s)$
in the following.
They are given by
\begin{align}
f_N(s)=\prod_{i=1}^N\frac{1}{1-s^i},
\label{cbi}
\end{align}
and we can easily confirm that these functions satisfy
(\ref{expansion2}) \cite{Gaiotto:2021xce}.

Next, let us discuss $f_{N,1}$.
It is convenient to define $g_{N,k}$ by
\begin{align}
g_{N,k}(s,u_i)=\frac{f_{N,k}(s,u_i)}{f_N(s)}.
\label{gdef}
\end{align}
We first focus on $g_{N,1}$,
and we denote them by $g_N$.
By picking up $\mathfrak{t}^1$ terms from (\ref{sdexpansion})
we obtain
\begin{align}
g_N(s,u_i)
-g_\infty(s,u_i)
=
\frac{f_\infty(s)}{f_N(s)}\sum_{m=0}^\infty s^{mN}f_m(s^{-1})g_m(s^{-1},\wt u_i).
\label{g1constraint}
\end{align}
By assumption, $f_N$ and $g_\infty$ are known functions and this equation gives
linear constraints on unknown functions $g_N$.

A nice property of $g_N$ ($N<\infty$) is that
they are fractional polynomials of $s$.
By ``a fractional polynomial of $s$''
we mean a function of $s$ in the form
\begin{align}
\sum_j c_j s^{r_j}
\end{align}
with a finite number of terms.
Unlike an ordinary polynomial, the exponents $r_j$ may not be non-negative integers.
They may be fractional and negative.
Because the number of terms is finite,
the exponents $r_j$ are
within a finite range depending on $N$.
This is not the case for $f_{N,1}$.
The $s$-expansion of $f_{N,1}$ includes infinitely many terms.
This is due to the presence of the Coulomb branch operators.
Let ${\cal O}$ be a BPS operator contributing to $f_{N,1}$.
Then, a product of ${\cal O}$ and Coulomb branch operators
also contributes to the same function $f_{N,1}$, and this makes $f_{N,1}$
an infinite series of $s$.
In the definition (\ref{gdef}) of $g_{N,k}$ the contribution from the
Coulomb branch operators is removed by the division by $f_N$.

To express the range of the exponents we use the notation
\begin{align}
s^{n^L(N)}\leq g_N(s,u_i)\leq s^{n^H(N)}.
\label{bound1}
\end{align}
This means that the exponents $r_j$ appearing in $g_N(s,u_i)$ are in the range
$n^L(N)\leq r_j\leq n^H(N)$.
If $g_N(s,u_i)$ is a fractional polynomial
bounded by (\ref{bound1}),
$g_N(s^{-1},\wt u_i)$ is also a fractional polynomial
bounded by
\begin{align}
s^{-n^H(N)}\leq g_N(s^{-1},\wt u_i)\leq s^{-n^L(N)}.
\label{bound2}
\end{align}
We do not need analytic continuation to relate $s$-expansions of $g_N(s,u_i)$ and $g_N(s^{-1},\wt u_i)$.

Let us consider inductive determination of $g_N$.
Namely, we suppose that we have obtained $g_N$ for $N\leq N_*-1$ with a fixed rank $N_*$,
and discuss whether we can determine $g_{N_*}$ by the relation (\ref{g1constraint}).
We can use (\ref{g1constraint}) in two different ways.
\begin{enumerate}
\item
If we set $N=N_*$ in (\ref{g1constraint}) the left hand side includes $g_{N_*}(s,u_i)$.
We obtain the relation
\begin{align}
g_{N_*}(s,u_i)
=
(\mbox{known})
+\frac{f_\infty(s)}{f_{N_*}(s)}\sum_{m=N_*}^\infty s^{mN_*}f_m(s^{-1})g_m(s^{-1},\wt u_i),
\label{stra1}
\end{align}
where ``$(\mbox{known})$'' includes known functions only.
Although the right hand side includes unknown functions $g_m(s^{-1},\wt u_i)$ with $m\geq N_*$,
if they give only higher order terms in $s$, then we may be able to extract information of
low order terms in $g_{N_*}(s,u_i)$.
\item
If we set $N$ in (\ref{g1constraint})
to be a value smaller than $N_*$, then the left hand side is
a known function, and the summand with $m=N_*$ on the right hand side includes $g_{N_*}(s^{-1},\wt u_i)$.
We can rewrite (\ref{g1constraint}) as
\begin{align}
g_{N_*}(s^{-1},\wt u_i)
=
(\mbox{known})
-\sum_{m=N_*+1}^\infty 
\frac{s^{mN}f_m(s^{-1})}{s^{N_*N}f_{N_*}(s^{-1})}
g_m(s^{-1},\wt u_i).
\label{g1constraint2}
\end{align}
With this relation it may be possible to extract information of low order terms in $g_{N_*}(s^{-1},\wt u_i)$,
or equivalently, high order terms in $g_{N_*}(s,u_i)$.
\end{enumerate}
Because these two approaches give information of different parts of $g_{N_*}(s,u_i)$,
it may be possible to completely determine $g_{N_*}(s,u_i)$ by combining these two.

Let $\mathfrak{o}_1(N_*)$ be
the order of the second term in (\ref{stra1}).
If we use the bound (\ref{bound2})
we obtain
\begin{align}
\mathfrak{o}_1(N_*):=
\min_{m\geq N_*}\left(mN_*+\frac{m(m+1)}{2}-n^H(m)\right).
\label{order1}
\end{align}
(The second term in the parentheses comes from $f_N(s^{-1})=(-1)^Ns^{\frac{N(N+1)}{2}}f_N(s)$.)
To determine the minimum value
(\ref{order1})
we need to know $n^H(m)$.
Although we do not have precise expression for $n^H(N)$ at present,
we will later see that numerical analysis suggests that $n^H(N)$ grows
as a linear function of $N$.
Then,
the minimum $\mathfrak{o}_1(N_*)$ always exists
and we can determine $s^{n<\mathfrak{o}_1(N_*)}$ terms in $g_{N_*}(s,u_i)$
with the relation (\ref{stra1}) .

The leading order of the unknown part of
(\ref{g1constraint2}) for each $m$ is
\begin{align}
(m-N_*)N+\frac{m(m+1)}{2}-\frac{N_*(N_*+1)}{2}-n^H(m).
\end{align}
$N$ is assumed to be smaller than $N_*$.
We want the order to be as large as possible.
To maximize the order we set $N=N_*-1$, and then
the order of the unknown part of (\ref{g1constraint2}), which we denote by $\mathfrak{o}_2(N_*)$,
becomes
\begin{align}
\mathfrak{o}_2(N_*):=
\min_{m\geq N_*+1}\left((m-N_*)\frac{m+3N_*-1}{2}-n^H(m)\right).
\label{order2}
\end{align}
Again, under the assumption of linear growth of $n^H(m)$,
the minimum always exists.
We can determine $s^{n<\mathfrak{o}_2(N_*)}$ terms in $g_{N_*}(s^{-1},\wt u_i)$,
or, equivalently, $s^{n>-\mathfrak{o}_2(N_*)}$ terms in $g_{N_*}(s,u_i)$.

Let us combine the two results.
The relation
(\ref{stra1}) determines $s^{n<\mathfrak{o}_1(N_*)}$ terms in $g_{N_*}(s,u_i)$,
and the relation
(\ref{g1constraint2}) determines $s^{n>-\mathfrak{o}_2(N_*)}$ terms in $g_{N_*}(s,u_i)$.
Therefore,
we can determine $g_{N_*}(s,u_i)$ completely
by combining these two relations if
\begin{align}
0<\mathfrak{o}_1(N_*)+\mathfrak{o}_2(N_*).
\label{g1bound}
\end{align}
Under the assumption of linear growth of $n^H(N)$,
this always hold for sufficiently large $N_*$.
Namely, there exists $N_{\rm crit}$ such that (\ref{g1bound}) holds for
arbitrary $N_*\geq N_{\rm crit}$.
The value of $N_{\rm crit}$ depends on the explicit form of $n^H(N)$.
If $N_{\rm crit}$ is not so large, we can determine $g_N$ for all $N$
by directly calculating $g_N$ for $N<N_{\rm crit}$ and apply the procedure
explained above.

The same strategy can be applied to the higher order terms in the expansion
(\ref{expansion2}).
To avoid the analysis becomes too abstract,
we first apply the above analysis for $g_N(s,u_i)$ to concrete examples
and then we will proceed to the analysis of the higher order terms.

\section{${\cal N}=2$ superconformal field theories}
\paragraph{Brane realization}
We consider the ${\cal N}=2$ superconformal theory realized on a stack of $N$ D3-branes probing a $7$-brane background
of the form
\begin{align}
\RR^{1,3}\times \RR^4_H\times {\cal C}.
\end{align}
The D3 branes extend along $\RR^{1,3}$, the four-dimensional Minkowski space.
They probe the six-dimensional transverse space $\RR^4_H\times{\cal C}$, where
$\RR^4_H=\CC_x\times \CC_y$ is the flat four-dimensional space and ${\cal C}$ is the
complex one-dimensional cone with the deficit angle $2\pi(1-\Delta^{-1})$.
If $N=1$ the transverse space ${\cal C}$ and $\RR_H^4$ are identified with the Coulomb branch and the Higgs branch
of the moduli space, respectively.
$\Delta$ is the scale dimension of the Coulomb branch operator with the smallest dimension.
The tip of ${\cal C}$ is singular and gives the $7$-brane worldvolume in the ten-dimensional spacetime.

Let $X$, $Y$, and $Z$ be complex coordinates of
$\CC_x$, $\CC_y$, and $\cal C$, respectively.
Due to the deficit angle,
$Z$ is restricted by
\begin{align}
0\leq \arg Z\leq \frac{2\pi}{\Delta}
\label{zrange}
\end{align}
and boundary points of the sector are identified by
\begin{align}
Z=r
\quad\sim\quad
Z=re^{\frac{2\pi i}{\Delta}}\quad(r\in\RR_{\geq0}).
\label{identification}
\end{align}

There exist seven non-trivial $\cal C$ specified by
the values of $\Delta$.
We use the symbols $G=H_0$, $H_1$, $H_2$, $D_4$, $E_6$, $E_7$, and $E_8$
to specify one of them.
See Table \ref{sevenbranes}.
\begin{table}[htb]
\caption{}\label{sevenbranes}
\centering
\begin{tabular}{ccccccccc}
\hline
\hline
$G$        & $H_0$         & $H_1$         & $H_2$         & $D_4$   & $E_6$         & $E_7$         & $E_8$ \\
\hline
$\Delta$ & $\frac{6}{5}$ & $\frac{4}{3}$ & $\frac{3}{2}$ & $2$     & $3$ & $4$ & $6$ \\
\hline
symmetry   & -             & $su(2)$       & $su(3)$       & $so(8)$ & $e_6$         & $e_7$         & $e_8$ \\
\hline
\end{tabular}
\end{table}
We also use $G$ to specify the gauge symmetry realized on the $7$-brane.
We denote the theory on $N$ D3-branes by $G[N]$.

Each of these theories includes a free hypermultiplet corresponding to the
center-of-mass motion of D3-branes along $\RR_H^4$.
Although these non-interacting degrees of freedom are often excluded from the
definition of the theory, we includes their contribution to the superconformal index
in the following analysis.
The rank $1$ theories $H_0[1]$, $H_1[1]$, and $H_2[1]$ with the free hypermultiplet excluded
are the Argyres-Douglas theories
\cite{Argyres:1995jj,Argyres:1995xn}
often called $(A_1,A_2)$, $(A_1,A_3)$, and $(A_1,D_4)$, respectively.
$E_n[1]$ ($n=6,7,8$) with the free hypermultiplet excluded are
called Minahan-Nemeschansky theories \cite{Minahan:1996fg,Minahan:1996cj}.

In addition to seven non-trivial cases, we regard ${\cal N}=4$ $U(N)$ SYM
as the special case with $\Delta=1$.

\paragraph{Superconformal index}
The global symmetry of $G[N]$ is
\begin{align}
su(2,2|2)\times su(2)_F\times G,
\end{align}
where $su(2,2|2)$ is the ${\cal N}=2$ superconformal algebra
with Cartan generators $H$, $J$, $\ol J$, $R$, and $r$.
$H$ is the Hamiltonian,
$J$ and $\ol J$ are the left and the right spins,
and $R$ and $r$ are $su(2)_R$ and $u(1)_r$ charges.
We normalize these generators so that the supercharges carry the following quantum numbers
\begin{align}
Q:(H,J,\ol J,R,r)&=(+\tfrac{1}{2},\pm\tfrac{1}{2},0,\pm\tfrac{1}{2},-\tfrac{1}{2}),\nonumber\\
\ol Q:(H,J,\ol J,R,r)&=(+\tfrac{1}{2},0,\pm\tfrac{1}{2},\pm\tfrac{1}{2},+\tfrac{1}{2}).
\end{align}
$su(2)_F\times G$ is the flavor symmetry,
and we use $F$ and $T_i$ ($i=1,\ldots,\rank G$) for their Cartan generators.
We normalize $F$ to be half integer.
$su(2)_F$ and $G$ are realized
as the isometry of $\RR_H^4$ and 
the gauge symmetry on the 7-brane, respectively.
For the ${\cal N}=4$ SYM $su(2,2|2)\times su(2)_F$ is enhanced to
$psu(2,2|4)$, and $G$ is trivial.

Let $\ol{\cal Q}$ be the component of $\ol Q$ with the quantum numbers
\begin{align}
\ol{\cal Q}:(H,J,\ol J,R,r)
=
(+\tfrac{1}{2},0,-\tfrac{1}{2},+\tfrac{1}{2},+\tfrac{1}{2}).
\end{align}
The superconformal index respecting the supercharge $\ol{\cal Q}$
is defined by
\begin{align}
I=\tr\left[(-1)^Ft^{2\ol J+2R}u^{2J}v^{2F}z^{\ol J+r}\prod_{i=1}^{\rank G} a_i^{T_i}\right],
\label{scidef}
\end{align}
where $t$, $u$, $v$, $z$, and $a_i$ are independent fugacities.

It is also convenient to introduce
generators $J_1$, $J_2$, $R_x$, $R_y$, and $R_z$ by
\begin{align}
J_1=\ol J+J,\quad
J_2=\ol J-J,\quad
R_x=R+F,\quad
R_y=R-F,\quad
R_z=r.
\end{align}
$R_x$, $R_y$, and $R_z$ rotate
$\CC_x$, $\CC_y$, and ${\cal C}$, respectively.
Corresponding to the new set of generators
we define
\begin{align}
q=tz^{1/2}u,\quad
p=tz^{1/2}u^{-1},\quad
x=tv,\quad
y=tv^{-1}.
\label{qpxy}
\end{align}
Four fugacities defined
in (\ref{qpxy}) and $z$ are not independent but constrained
by $qp=xyz$.
With these variables, the index
(\ref{scidef}) is rewritten as
\begin{align}
I=\tr\left[(-1)^Fq^{J_1}p^{J_2}x^{R_x}y^{R_y}z^{R_z}
\prod_{i=1}^{\rank G} a_i^{T_i}\right].
\label{defindex2}
\end{align}
Only states saturating the BPS bound
\begin{align}
0\leq
\{\ol{\cal Q},\ol{\cal Q}^\dagger\}
&=H-2\ol J-2R-r
\nonumber\\
&=H-J_1-J_2-R_x-R_y-R_z
\label{bpsqbar}
\end{align}
contribute to the index.

\paragraph{Holographic dual}

The holographic dual geometry is $AdS_5\times S^5_\Delta$, where
$S^5_\Delta$ is a five-dimensional sphere with a conical defect along a large $S^3$
\cite{Fayyazuddin:1998fb,Aharony:1998xz}.
$S^5_\Delta$ is explicitly defined by
\begin{align}
|X|^2+|Y|^2+|Z|^2=1^2,
\end{align}
with the coordinates introduced above.

Under the assumption that giant graviton contributions
localize at the $U(1)_{R_x}\times U(1)_{R_y}\times U(1)_{R_z}$ fixed points in the D3-brane configuration space,
we expect that giants wrapped on the following
three three-cycles in $S_\Delta^5$ contribute to the index,
\begin{align}
X=0,\quad
Y=0,\quad
Z=0,
\end{align}
and the corresponding giant graviton expansion is \cite{Imamura:2021dya}
\begin{align}
\frac{I_N}{I_\infty}=\sum_{m_x,m_y,m_z=0}^\infty x^{m_xN}y^{m_yN}z^{\Delta m_zN}F_{m_x,m_y,m_z},
\label{ggez}
\end{align}
where $F_{m_x,m_y,m_z}$ are the indices of the theories
realized on the giant gravitons
with wrapping numbers $m_x$, $m_y$, and $m_z$.
This triple-sum expansion were studied in \cite{Imamura:2021dya}
with a focus on the contributions from $(m_x,m_y,m_z)=(1,0,0)$ and $(0,1,0)$.

\paragraph{Self-dual giant graviton expansion}

We are interested in a self-dual giant graviton expansion.
This requires the theory on giant gravitons $T_N'$ are
the same as the original theory $T_N$.
For this to be the case the giant gravitons must wrap on the singular locus $Z=0$.
This is the case if we choose expansion variable $\mathfrak{t}=t$.
(We will later slightly change the definition of $\mathfrak{t}$.
See (\ref{tsu}).)

With this choice of the expansion variable,
two cycles $X=0$ and $Y=0$ decouple \cite{Imamura:2022aua,Fujiwara:2023bdc}.
We obtain the giant graviton expansion of the form
\begin{align}
\frac{I_N}{I_\infty}=\sum_{m=0}^\infty z^{m\Delta_G}F_{0,0,m},
\end{align}
where $F_{0,0,m}$ is essentially the same as $I_m$.
The variable change transforming $F_{0,0,m}$ into $I_m$ is related
to an outer automorphism of the unbroken symmetry on
the worldvolume of giant gravitons \cite{Arai:2019xmp}.
The symmetry preserved by giant gravitons wrapped around the cycle $Z=0$ is
\begin{align}
u(2|2)\times su(2)_J\times su(2)_F\times G,
\label{symalg}
\end{align}
where $u(2|2)=su(2|2)\rtimes u(1)_r$ is the algebra generated by $\ol Q$ and $\ol Q^\dagger$.
(More precisely, $\ol Q$ and $\ol Q^\dagger$ generates $su(2|2)$, and the extension by its outer automorphism $u(1)_r$
generated by $r=R_z$ gives $u(2|2)$.)
The outer automorphism of the preserved symmetry algebra
(\ref{symalg})
\begin{align}
\ol J\leftrightarrow R,\quad
J\leftrightarrow F,\quad
R_z\rightarrow -R_z,\quad
H-R_z\rightarrow H-R_z
\label{auto1}
\end{align}
relates the theory on the boundary and the theory on the cycle $Z=0$.
This is consistent with the definition of the superconformal index (\ref{defindex2}),
and the definition is invariant under
(\ref{auto1}) if we do the following variable change $\sigma_z$ at the same time.
\begin{align}
\sigma_z(q,p,x,y,z)=(x,y,q,p,z^{-1}),\quad
\sigma_z(t,z,y,u)
=(tz^{1/2},z^{-1},u,y)
\label{sigmaz1}
\end{align}
To match the form of the variable change with
the one used in previous sections, we define the following variables.
\begin{align}
\mathfrak{t}&=tz^{\frac{1}{4}},\nonumber\\
s&=z^\Delta,\nonumber\\
u_i&=(u,v,a_i),\nonumber\\
\wt u_i&=(v,u,a_i).
\label{tsu}
\end{align}
Then, $\sigma_z$ acts on these variables as follows.
\begin{align}
\sigma_z(\mathfrak{t},s,u_i)
=(\mathfrak{t},s^{-1},\wt u_i),
\end{align}
and (\ref{ggez}) is rewritten into the form (\ref{sdexpansion}).

\paragraph{Schur index}
The Schur index
\cite{Gadde:2011uv}
is defined by tuning the fugacities so that
another supercharge ${\cal Q}$ in $Q$
is respected.
Without loosing generality we can choose ${\cal Q}$ with
the quantum numbers
\begin{align}
{\cal Q}:(H,J,\ol J,R,r)
=(+\tfrac{1}{2},-\tfrac{1}{2},0,+\tfrac{1}{2},-\tfrac{1}{2}).
\end{align}
To respect this supercharge we need to set
$p=z$.

The Schur index is in general more tractable than
the general superconformal index, and it is often possible to
obtain analytic expression for them \cite{Bourdier:2015wda,Bourdier:2015sga,Pan:2021mrw,Hatsuda:2022xdv}.
It would be nice if we could discuss self-dual giant graviton expansion
for Schur index.
Unfortunately, it is not possible.
The BPS bound associated with ${\cal Q}$ is
\begin{align}
0\leq
\{{\cal Q},{\cal Q}^\dagger\}
&=H-2J-2R+r\nonumber\\
&=H-J_1+J_2-R_x-R_y+R_z.
\label{bpsq}
\end{align}
Only states saturating two bounds
(\ref{bpsqbar}) and (\ref{bpsq})
at the same time contribute to the Schur index,
but giant gravitons wrapped on $Z=0$
does not saturate (\ref{bpsq}).
Furthermore, the Schur limit $p=z$ is
inconsistent with the variable change (\ref{sigmaz1}).
For these reasons
we will not discuss the Schur limit
in this work.

\paragraph{Large $N$ limit}

The relation between the operator spectrum of $G[\infty]$ and Kaluza-Klein modes in
the dual geometry was investigated in \cite{Fayyazuddin:1998fb,Aharony:1998xz},
and the corresponding index was calculated in \cite{Imamura:2021dya}.

Let $i_\infty$ be the letter index in the holographic description.
The large $N$ index $I_\infty$ is given by $I_\infty=\Pexp i_\infty$.
$i_\infty$ is the sum of two contributions \cite{Imamura:2021dya}:
\begin{align}
i_\infty=f^{\rm(sugra)}+f^{(7)}\chi_{\rm adj}^G.
\label{iinfty}
\end{align}
$f^{\rm(sugra)}$ is the letter index of the supergravity
multiplet in the ten-dimensional bulk
\begin{align}
f^{\rm(sugra)}
&=\frac{[(x+y)-xy+xy(q+p)]-[(x'+y')-x'y'+x'y'(q'+p')]}{(1-q)(1-p)(1-x)(1-y)}
\nonumber\\
&+\frac{s}{1-s}\frac{(1-q')(1-p')(1-x')(1-y')}{(1-q)(1-p)(1-x)(1-y)},
\label{fsugra}
\end{align}
where we defined
\begin{align}
q'=\frac{q}{z},\quad
p'=\frac{p}{z},\quad
x'=zx,\quad
y'=zy.
\end{align}
The second term in (\ref{iinfty}) is the
contribution from the vector multiplet on
the seven-brane.
$\chi_{\rm adj}^G$ is the adjoint character of the symmetry $G$
and
$f^{(7)}$ is the letter index of the $u(1)$ vector multiplet on the eight-dimensional
worldvolume of the seven-brane
\begin{align}
f^{(7)}=\frac{xy-pq}{(1-q)(1-p)(1-x)(1-y)}.
\end{align}
We can write the letter index for $\Delta=1$ (corresponding to the ${\cal N}=4$ $U(N)$ SYM) in the form
(\ref{iinfty}) by formally setting $\chi_{\rm adj}^G=-2$.

\section{Observation}
\begin{figure}[htb]
\centering
\includegraphics{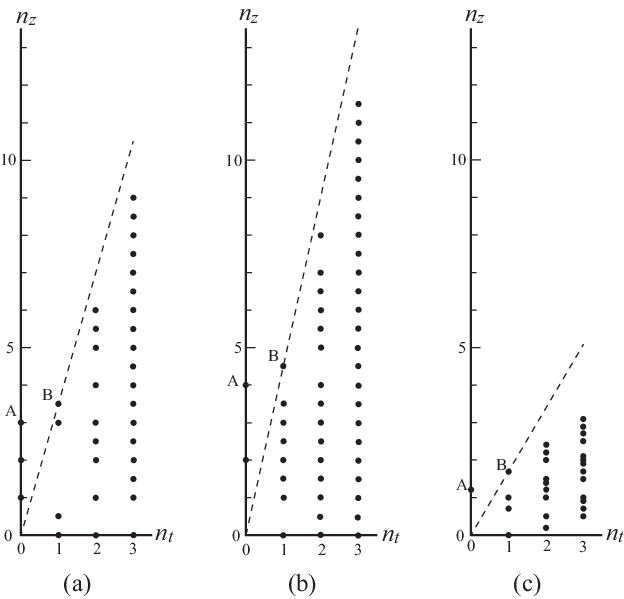}
\caption{2d plots of $\Plog I$. (a) ${\cal N}=4$ $U(3)$ SYM (b) ${\cal N}=2$ $Sp(2)$ SQCD ($D_4[2]$) (c) $H_0[1]$.
$A$ is the Coulomb branch operator with the largest dimension and $B$ is its conformal descendant determining the envelope line.}\label{n4plot.eps}
\end{figure}

The arguments in Section \ref{strategy} is based on the assumption of linear growth of $n^H(N)$.
To confirm this and to obtain more explicit form of $n^H(N)$,
let us look at some examples obtained by numerical calculations.
We use two-dimensional plots to express the structure of the series expansion
of indices.
See Figure \ref{n4plot.eps}.
We showed the plethystic logarithm of the index, $\Plog I$, as a two-dimensional plot
for three theories: ${\cal N}=4$ $U(3)$ SYM ($\Delta=1$, $N=3$), $D_4[2]$ ($\Delta=2$, $N=2$), and $H_0[1]$ ($\Delta=\frac{6}{5}$, $N=1$).
The horizontal and vertical axes are the exponents of $t$ and $z$, respectively.
Namely, a term $f(u,v)t^{n_t}z^{n_z}$ is shown as a dot at coordinates $(n_t,n_z)$.

Due to the BPS bounds all dots are in the first quadrant (including its boundary) in the $n_t$-$n_z$ plane,
and $n_t$ is quantized to be integer.
There are $N$ dots at even intervals on the vertical axis.
They correspond to the Coulomb branch operators $\varphi_k$ $(k=1,\ldots,N)$ with
dimensions $k\Delta$.

Except for those on the vertical axis, there are no dots above the
envelope line passing through the origin and the dot at $(1,N\Delta+\frac{1}{2})$,
which is labeled by $B$ in Figure \ref{n4plot.eps}.
The dot $B$ corresponds to conformal descendants $\partial \varphi_N$ of the
Coulomb branch operator $\varphi_N$ with the largest dimension.
This means that
the exponents $n_t$ and $n_z$ satisfy
\begin{align}
0\leq n_z\leq n_t\left(N\Delta+\frac{1}{2}\right).
\label{boundtz}
\end{align}
We rewrite the index as the function of $\mathfrak{t}$, $s$, and $u_i$,
and express it as $\mathfrak{t}$ series
\begin{align}
I_N(\mathfrak{t},s,u_i)=f_N(s)\left[1+\sum_{k=1}^\infty \mathfrak{t}^k g_{N,k}(s,u_i)\right].
\label{iNexpansion}
\end{align}
The bound (\ref{boundtz}) means
\begin{align}
s^{n^L_k(N)}
\leq g_{N,k}(s,u_i)\leq
s^{n^H_k(N)}.
\label{bound45}
\end{align}
with
\begin{align}
n^L_k=-\frac{k}{4\Delta},\quad
n^H_k=k\left(N+\frac{1}{4\Delta}\right).
\label{nlnh}
\end{align}
Although we have no rigorous proof of (\ref{nlnh}),
let us take it as a working hypothesis.

\section{$\mathfrak{t}^1$ terms}
Let us continue the analysis of $g_N(=g_{N,1})$ in Section \ref{strategy} by using
the upper bound $n^H(N)=n^H_1(N)$:
\begin{align}
n^H(N)=N+\frac{1}{4\Delta}.
\end{align}
By substituting this into 
(\ref{order1})
and (\ref{order2})
we obtain
\begin{align}
\mathfrak{o}_1(N_*)&=\frac{3}{2}N_*^2-\frac{1}{2}N_*-\frac{1}{4\Delta},\label{o1k1}\\
\mathfrak{o}_2(N_*)&=N_*-1-\frac{1}{4\Delta},\label{o2k1}
\end{align}
for positive $N_*$,
and the condition (\ref{g1bound}) becomes
\begin{align}
\frac{3}{2}N_*^2+\frac{1}{2}N_*-1-\frac{1}{2\Delta}>0.
\end{align}
This condition is satisfied for all positive $N_*$.
Namely, if we know the index for $N=0$ (and $N=\infty$),
we can uniquely determine $g_N$ for all $N$ by requiring
the self-dual giant graviton expansion holds.
The theory with $N=0$ is the trivial theory with $I_0=1$,
and $g_0=0$.

We can give an analytic solution satisfying the relation
(\ref{g1constraint}).
Let us take the ansatz
\begin{align}
g_N(s,u_i)=(1-s^N)g_\infty(s,u_i).
\label{ansatzgn}
\end{align}
This is consistent with the large $N$ limit, and correctly reproduces $g_0=0$.
Although we can obtain $g_\infty$ 
by expanding (\ref{iinfty}), let us discuss without using the explicit form for the present.
By substituting
(\ref{ansatzgn}) into
(\ref{g1constraint})
the right hand side
of (\ref{g1constraint})
becomes
\begin{align}
(\mbox{r.h.s})
&=\frac{f_\infty(s)}{f_N(s)}\sum_{m=1}^\infty s^{mN}f_{m-1}(s^{-1})g_\infty(s^{-1},\wt u_i)
\nonumber\\
&=s^N\frac{f_\infty(s)}{f_N(s)}\sum_{m=0}^\infty s^{mN}f_m(s^{-1})g_\infty(s^{-1},\wt u_i)
\nonumber\\
&=s^Ng_\infty(s^{-1},\wt u_i)
\end{align}
At the second equality $m$ is replaced by $m+1$
and at the third equality (\ref{expansion2}) is used.
The left hand side
of (\ref{g1constraint})
becomes $-s^Ng_\infty(s,u_i)$,
and the relation
(\ref{g1constraint}) reduces to
\begin{align}
g_\infty(s^{-1},\wt u_i)=-g_\infty(s,u_i).
\label{oddh}
\end{align}
The consistency requires $g_\infty(s,u_i)$ satisfy
the non-trivial relation (\ref{oddh}).

In fact, this is the case.
As is pointed out in \cite{Fujiwara:2023bdc},
the large $N$ letter indices $i_\infty$
satisfies the interesting relation
\begin{align}
i_\infty(\mathfrak{t},s,u_i)
+i_\infty(\mathfrak{t},s^{-1},\wt u_i)=-1,
\label{selfduali}
\end{align}
and the $\mathfrak{t}^1$ terms in
(\ref{selfduali}) give (\ref{oddh}).

(\ref{oddh}) can be also confirmed with
the explicit form of $g_\infty(s,u_i)$
obtained by expanding (\ref{iinfty});
\begin{align}
g_\infty
=\frac{1-z}{1-s}z^{-\frac{1}{4}}(\chi_1^v-sz^{-\frac{1}{2}}\chi_1^u),
\label{binf}
\end{align}
where $\chi_n^u$ and $\chi_n^v$ are $su(2)_J$ and $su(2)_F$ characters,
respectively;
\begin{align}
\chi_n^u=\frac{u^{n+1}-u^{-n-1}}{u-u^{-1}},\quad
\chi_n^v=\frac{v^{n+1}-v^{-n-1}}{v-v^{-1}}.
\end{align}

By substituting (\ref{binf}) into
(\ref{ansatzgn}) we obtain $g_N$;
\begin{align}
g_N(s,u_i)
&=\frac{(1-z)(1-s^N)}{1-s}z^{-\frac{1}{4}}[\chi_1^v-sz^{-\frac{1}{2}}\chi_1^u].
\label{gn1}
\end{align}
Because the prescription in Section \ref{strategy} unambiguously
determine $g_N$, (\ref{ansatzgn}) is the unique solution
to (\ref{g1constraint}).
Indeed, we can confirm that
(\ref{gn1}) is consistent with the known results for $N=1$.
The superconformal indices of
the rank $1$ theories, $H_0[1]$ \cite{Maruyoshi:2016tqk},
$H_1[1]$ \cite{Maruyoshi:2016aim},
$H_2[1]$ \cite{Agarwal:2016pjo}, and $D_4[1]$
are known and given by the unified form
\begin{align}
I_1
&=\Pexp\bigg[s
+(1-z)(-sz^{-\frac{1}{2}}\chi_1^u+\chi_1^v)t
\nonumber\\
&+(1-z)(
sz^{-1}
-z
-s^2z^{-1}
-s\chi_2^u
+z^{\frac{1}{2}}\chi_1^u\chi_1^v
+(1-s)\chi_{\rm adj}^G
)t^2
+{\cal O}(t^3)\bigg].
\label{ione}
\end{align}
up to $t^2$ terms (See Appendix).
The indices of $E_6[1]$
\cite{Gadde:2010te}
and $E_7[1]$
\cite{Agarwal:2018ejn}
are also known, and
(\ref{ione}) is consistent with the terms explicitly shown in the references.
(\ref{ione}) also gives the index of the ${\cal N}=4$ $U(1)$ super Maxwell theory
by setting $\Delta=1$ and $\chi_{\rm adj}^G=-2$.
(\ref{gn1}) agrees with $g_1$ extracted from
(\ref{ione}).

\section{$\mathfrak{t}^{k\geq 2}$ terms}
Let us proceed to higher order terms
in the expansion (\ref{texpansion}).
We discuss
whether we can determine $g_{N_*,k_*}$ for fixed $N_*$ and $k_*$
when we know $g_{N,k}$ for $k\leq k_*-1$
and $g_{N,k_*}$ for $N\leq N_*-1$.
Basic strategy is the same as before.
We use the giant graviton expansion in two ways.

First, we
extract $\mathfrak{t}^{k_*}$ terms from the giant graviton expansion of $I_{N_*}$.
We obtain
\begin{align}
g_{N_*,k_*}(s,u_i)
&=(\mbox{known})+\frac{f_\infty(s)}{f_{N_*}(s)}
\sum_{m=N_*}^\infty s^{mN_*}f_m(s^{-1})g_{m,k_*}(s^{-1},\wt u_i).
\end{align}
This is essentially the same as (\ref{stra1}),
and we obtain the following order of the unknown part.
\begin{align}
\mathfrak{o}_{1,k_*}(N):=\min_{m\geq N_*}\left(mN_*+\frac{m(m+1)}{2}-n^H_{k_*}(m)\right).
\end{align}
Only difference from (\ref{order1}) is
that $n^H(m)$ is replaced by $n^H_{k_*}(m)$.
By using $n^H_{k_*}$ in (\ref{nlnh})
we obtain
\begin{align}
\mathfrak{o}_{1,k_*}(N)=\frac{3}{2}N_*^2+\frac{1}{2}N_*-k_*\left(N_*+\frac{k_*}{4\Delta}\right).
\end{align}

Another relation is obtained from ${\cal O}(\mathfrak{t}^{k_*})$ terms
in the giant graviton expansion of $I_N$ with $N<N_*$.
We obtain the relation corresponding to (\ref{g1constraint2})
\begin{align}
g_{N_*,k_*}(s^{-1},\wt u_i)
=
(\mbox{known})
-\sum_{m=N_*+1}^\infty
\frac{s^{m N}f_m(s^{-1})}{s^{N_*N}f_{N_*}(s^{-1})}
g_{m,k_*}(s^{-1},\wt u_i).
\end{align}
To make the order of the unknown part as large as possible
we set $N=N_*-1$ and obtain
\begin{align}
\mathfrak{o}_{2,k_*}(N)
&=\min_{m\geq N_*+1}\left((m-N_*)N+\frac{m(m+1)}{2}-\frac{N_*(N_*+1)}{2}-n^H_k(m)\right)
\nonumber\\
&=2N_*-k\left(N_*+1+\frac{1}{4\Delta}\right).
\end{align}
Therefore, the condition that we can completely determine $g_{N,k}$ is
\begin{align}
0<\mathfrak{o}_{1,k_*}(N)+\mathfrak{o}_{2,k_*}(N)
=
\frac{3}{2}N_*^2+\frac{5}{2}N_*-k_*\left(2N_*+1+\frac{1}{2\Delta}\right).
\label{o1o2ineq}
\end{align}
Again, this condition is satisfied for sufficiently large $N_*$.

For $k_*=2$,
this is satisfied for $N_*\geq N_{\rm crit}=2$ (for $\Delta>1$) or for $N_*\geq N_{\rm crit}=3$ (for $\Delta=1$).
Let us consider $\Delta>1$ case first.
Because $N_{\rm crit}=2$, it is guaranteed that
if we use $g_{\infty,2}$ extracted from
(\ref{iinfty}),
$g_{1,2}$ extracted from
(\ref{ione}), and $g_{0,2}=0$ as
input data, the self-dual giant graviton expansion
uniquely determines $g_{N,2}$ for arbitrary $N$.
We find the following solution.
\begin{align}
g_{N,2}(s,u_i)&=\frac{(1-z)(1-s^N)}{1-s}z^{-\frac{1}{2}}(
sz^{-1}(1-s^N)
-z
+(z^{\frac{1}{2}}-(s-s^N)z^{-\frac{1}{2}})\chi_1^u\chi_1^v
\nonumber\\&
-s\chi_2^u+(1-s^{N-1})\chi_2^v
+(1-s)\chi_{\rm adj}^G)
+\frac{1}{2}(g_{N,1}(s^2,u_i^2)+g_{N,1}^2(s,u_i)).
\label{gn2}
\end{align}
By combining $g_{N,1}(=g_N)$ in (\ref{gn1})
and 
$g_{N,2}$ in (\ref{gn2}), we obtain the index
\begin{align}
I_N
&=\Pexp\bigg[
\frac{s(1-s^N)}{1-s}+\frac{(1-z)(1-s^N)}{1-s}(-sz^{-\frac{1}{2}}\chi_1^u+\chi_1^v)t
\nonumber\\
&+\frac{(1-z)(1-s^N)}{1-s}(
sz^{-1}(1-s^N)
-z
+(z^{\frac{1}{2}}-(s-s^N)z^{-\frac{1}{2}})\chi_1^u\chi_1^v
\nonumber\\&
-s\chi_2^u+(1-s^{N-1})\chi_2^v
+(1-s)\chi_{\rm adj}^G
)t^2+{\cal O}(t^3)\bigg]
\end{align}
This is the main result of this paper.
We can confirm that this correctly reproduces the index of ${\cal N}=4$ $U(N)$ SYM (with $\Delta=1$), too.

We conclude this work by a short comment on $\mathfrak{t}^{\geq3}$ terms.
The strategy will work for higher order terms
as long as we can prepare input data for $N<N_{\rm crit}$.
For example, for $k_*=3$,
(\ref{o1o2ineq}) gives the inequality
\begin{align}
6+\frac{3}{\Delta}<3N_*^2-7N_*,
\end{align}
and regardless of the value of $\Delta$
this condition is satisfied when
$N_*\geq N_{\rm crit}=4$.
This means we need to prepare
$g_{3,N}$ with $N\leq 3$ as input data to start the induction procedure.
Unfortunately, as far as we know, there are no results for $g_{3,N}$ of $H_n[N]$ and $E_n[N]$ with
$N=2,3$ in the literature,
and we need some extra information to calculate higher order terms.

\section*{Acknowledgments}
The work of Y.~I. was supported by JSPS KAKENHI Grant Number JP21K03569.

\appendix
\section{Results for small rank theories}
In this appendix we will show some explicit results
for small rank theories.
We will show the plethystic logarithms of the indices
rather than the indices themselves for convenience.

We can calculate the indices of ${\cal N}=4$ $U(N)$ SYM
$I_{{\cal N}=4[N]}$ by using the localization formula.
Their plethystic logarithms for $N=1,2,3$ are
\begin{align}
\Plog I_{{\cal N}=4[1]}
&=z
+(1-z)(-z^{\frac{1}{2}}\chi_1^u+\chi_1^v)t
\nonumber\\
&+(1-z)(-1-z\chi_2^u+z^{\frac{1}{2}}\chi_1^u\chi_1^v)t^2
+{\cal O}(t^3),
\nonumber\\
\Plog I_{{\cal N}=4[2]}
&=z+z^2
+(1-z^2)(-z^{\frac{1}{2}}\chi_1^u+\chi_1^v)t
\nonumber\\
&+(1-z^2)(-1+z-z^2-z\chi_2^u+z^{\frac{3}{2}}\chi_1^u\chi_1^v+(1-z)\chi_2^v)t^2
+{\cal O}(t^3)
\nonumber\\
\Plog I_{{\cal N}=4[3]}
&=z+z^2+z^3
+(1-z^3)(-z^{\frac{1}{2}}\chi_1^u+\chi_1^v)t
\nonumber\\
&+(1-z^3)(-1+z-z^3-z\chi_2^u
+z^{\frac{5}{2}}\chi_1^u\chi_1^v
+(1-z^2)\chi_2^v)t^2
+{\cal O}(t^3)
\end{align}
$D_4[N]$ are also Lagrangian theories and we can easily obtain
\begin{align}
\Plog I_{D_4[1]}
&=z^2
+(1-z)(-z^{\frac{3}{2}}\chi_1^u+\chi_1^v)t
\nonumber\\
&+(1-z)(
-z^3
-z^2\chi_2^u
+z^{\frac{1}{2}}\chi_1^u\chi_1^v
+(1-z^2)\chi^{so(8)}_{\rm adj}
)t^2
\nonumber\\
&+{\cal O}(t^3),
\nonumber\\
\Plog I_{D_4[2]}
&=z^2+z^4+(1-z)(1+z^2)(-z^{\frac{3}{2}}\chi_1^u+\chi_1^v)t
\nonumber\\
&+(1-z)(1+z^2)(
-z^5
-z^2\chi_2^u
+(1-z^2)\chi_2^v
+(z^{\frac{1}{2}}-z^{\frac{3}{2}}+z^{\frac{7}{2}})\chi_1^u\chi_1^v
+(1-z^2)\chi_{\rm adj}^{so(8)}
)t^2
\nonumber\\
&+{\cal O}(t^3)
\end{align}
$H_0[1]$, $H_1[1]$, and $H_2[1]$ are Argyres-Douglas theories (with an extra free hypermultiplet),
which are often called
$(A_1,A_2)$, $(A_1,A_3)$, and $(A_1,D_4)$, respectively.
They are realized as infra-red RG fixed points of Lagrangian theories
\cite{Maruyoshi:2016tqk,Maruyoshi:2016aim,Agarwal:2016pjo},
and we can calculate the indices with the help of the $a$-maximization procedure \cite{Intriligator:2003jj}.
The results are
\begin{align}
\Plog I_{H_0[1]}
&=z^{\frac{6}{5}}
+(1-z)(-z^{\frac{7}{10}}\chi_1^u+\chi_1^v)t
\nonumber\\
&+(1-z)(
z^{\frac{1}{5}}
-z
-z^{\frac{7}{5}}
-z^{\frac{6}{5}}\chi_2^u
+z^{\frac{1}{2}}\chi_1^u\chi_1^v)t^2
\nonumber\\
&+{\cal O}(t^3),
\nonumber\\
\Plog I_{H_1[1]}
&=z^{\frac{4}{3}}
+(1-z)(-z^{\frac{5}{6}}\chi_1^u+\chi_1^v)t
\nonumber\\
&+(1-z)(
z^{\frac{1}{3}}
-z
-z^{\frac{5}{3}}
-z^{\frac{4}{3}}\chi_2^u
+z^{\frac{1}{2}}\chi_1^u\chi_1^v
+(1-z^{\frac{4}{3}})\chi_{\rm adj}^{su(2)}
)t^2
\nonumber\\
&+{\cal O}(t^3),
\nonumber\\
\Plog I_{H_2[1]}
&=z^{\frac{3}{2}}
+(1-z)(-z\chi_1^u+\chi_1^v)t
\nonumber\\
&+(1-z)(
z^{\frac{1}{2}}
-z
-z^2
-z^{\frac{3}{2}}\chi_2^u
+z^{\frac{1}{2}}\chi_1^u\chi_1^v
+(1-z^{\frac{3}{2}})\chi_{\rm adj}^{su(3)}
)t^2
\nonumber\\
&+{\cal O}(t^3).
\end{align}


\end{document}